\documentclass[prd,showpacs]{revtex4}
\begin{document}

\newcommand{\re}{\mathop{\mathrm{Re}}}

\newcommand{\be}{\begin{equation}}
\newcommand{\ee}{\end{equation}}
\newcommand{\bea}{\begin{eqnarray}}
\newcommand{\eea}{\end{eqnarray}}
\addtolength{\topmargin}{20mm}

\title{New approach to study gravitational stability of the solutions to the Einstein equations}

\author{Janusz Garecki}
\email{garecki@wmf.univ.szczecin.pl}
\affiliation{\it Institute of Mathematics University of Szczecin
and Cosmology Group University of Szczecin,
 Wielkopolska 15, 70-451 Szczecin, Poland}
\date{\today}
\input epsf
\begin{abstract}
Here we propose a new method to study gravitational stability of the solutions
to the Einstein equations. This method uses the canonical
superenergy tensors which have been introduced in past in our
papers and is very alike to the procedure of finding the stable
minima of the interior energy $U$ for a thermodynamical system.
\end{abstract}
\maketitle
\section{Introduction}
In the paper we propose a new approach to study gravitational stability of a solution
to the Einstein equations.
This approach uses the canonical superenergy tensors which were
introduced into general relativity in our papers \cite {Gar1}.
Namely, we assert that when the total superenergy density, matter
and gravitation, $\epsilon_s$, is non-negative, i.e., when $\epsilon_s\geq
0$, then the solution can be stable under small metric perturbation.
Contrary, when $\epsilon_s$ is negative-definite, i.e., when $\epsilon_s<0$,
then the solution cannot be gravitationally stable.

The paper is organized as follows. In Section II we remind problems
with local energy-momentum in general relativity and our proposition to
avoid them -- the canonical superenergy tensors.

In Section III we give examples of an intriguing correlation between
gravitational stability of the very known solutions to the Einstein equations
and sign of the total canonical superenergy density, $\epsilon_s$,
for them.
 We claim there that these exciting correlations are consequences of the Proposition,
which we have formulated and proved in this Section. From this Proposition there follows
the our above mentioned statement concerning stability.

Finally, the short Section IV contains our final remarks.

In Appendix we present in more details some results of the last our calculations.

In the paper we use the same signature and notation as used in the
last editions of the famous book by Landau and Lifshitz \cite{LL}.

The $\Lambda$ term which we consider in Section III and in
Appendix we treat as source term in Einstein equations, i.e., as
energy-momentum tensor of the form $_{\Lambda} T_i^{~k}= (-){\Lambda\over\beta}\delta_i^k
$.
\section{The canonical superenergy tensors}
In the framework of general relativity ({\bf GR}), as a
consequence of the Einstein Equivalence Principle ({\bf EEP}), the
gravitational field {\it has non-tensorial strengths} $\Gamma^i_{kl}
 = \{^i_{kl}\}$ and {\it admits no energy-momentum tensor}. One
 can only attribute to this field {\it gravitational
 energy-momentum pseudotensors}. The leading object of such a kind
 is the {\it canonical gravitational energy-momentum pseodotensor}
 $_E t_i^{~k}$ proposed already in past by Einstein. This
 pseudotensor is a part of the {\it canonical energy-momentum
 complex} $_E K_i^{~k}$ in {\bf GR}.

The canonical complex $_E K_i^{~k}$ can be easily obtained by
rewiriting Einstein equations to the superpotential form
\begin{equation}
_E K_i^{~k} := \sqrt{\vert g\vert}\bigl( T_i^{~k} + _E
t_i^{~k}\bigr) = _F U_i^{~[kl]}{}_{,l}
\end{equation}
where $T^{ik} = T^{ki}$ is the symmetric energy-momentum tensor for matter, $g = det[g_{ik}]$,
 and

\begin{eqnarray}
_E t_i^{~k}& =& {c^4\over 16\pi G} \bigl\{\delta_i^k
g^{ms}\bigl(\Gamma^l_{mr}\Gamma^r_{sl} -
\Gamma^r_{ms}\Gamma^l_{rl}\bigr)\nonumber\cr
&+& g^{ms}_{~~,i}\bigl[\Gamma^k_{ms} - {1\over 2}
\bigl(\Gamma^k_{tp}g^{tp} -
\Gamma^l_{tl}g^{kt}\bigr)g_{ms}\nonumber\cr
&-& {1\over 2}\bigl(\delta^k_s \Gamma^l_{ml} +
\delta^k_m \Gamma^l_{sl}\bigr)\bigr]\bigr\};
\end{eqnarray}
\begin{equation}
_F {U_i^{~[kl]}} = {c^4\over 16\pi G}g_{ia}({\sqrt{\vert
g\vert}})^{(-1)}\bigl[\bigl(-g\bigr)\bigl(g^{ka} g^{lb} - g^{la}
g^{kb}\bigr)\bigr]_{,b}.
\end{equation}
$_E t_i^{~k}$ are components of the canonical energy-momentum
pseudotensor for gravitational field $\Gamma ^i_{kl} =
\bigl\{^i_{kl}\bigr\}$, and $_F {U_i^{~[kl]}}$ are von Freud
superpotentials.
\begin{equation}
_E K_i^{~k} = \sqrt{\vert g\vert}\bigl(T_i^{~k} + _E
t_i^{~k}\bigr)
\end{equation}
are components of the {\it Einstein canonical energy-momentu complex,
for matter and gravity}, in {\bf GR}.

In consequence of (1) the complex $_E K_i^{~k}$satisfies local
conservation laws
\begin{equation}
{_E K_i^{~k}}_{,k}\equiv 0.
\end{equation}
In very special cases one can obtain from these local conservation
laws the reasonable integral conservation laws.

Despite that one can easily introduce in {\bf GR} {\it the
canonical (and others) superenergy tensor} for gravitational
field. This was done in past in a series of our articles (See,
e.g.,\cite{Gar1} and references therein).
It appeared that the idea of the superenergy tensors is universal:
to any physical field having an energy-momentum tensor or
pseudotensor one can attribute the coresponding superenergy
tensor.

So, let us give a short reminder of the general, constructive
definition of the superenergy tensor $S_a^{~b}$ applicable to
gravitational field and to any matter field. The definition uses
{\it locally Minkowskian structure} of the spacetime in {\bf GR}
and, therefore, it fails in a spacetime with torsion, e.g., in Riemann-Cartan
spacetime.

In normal Riemann coordinates {\bf NRC(P)} we define (pointwiese)
\begin{equation}
S_{(a)}^{~~~(b)}(P) = S_a^{~b} :=(-) \displaystyle\lim_{\Omega\to
P}{\int\limits_{\Omega}\biggl[T_{(a)}^{~~~(b)}(y) - T_{(a)}^{
~~~(b)} (P)\biggr]d\Omega\over 1/2\int\limits_{\Omega}\sigma(P;y)
d\Omega},
\end{equation}
where
\begin{eqnarray}
T_{(a)}^{~~~(b)}(y) &:=& T_i^{~k}(y)e^i_{~(a)}(y)
e_k^{~(b)}(y),\nonumber\cr
T_{(a)}^{~~~(b)}(P)&:=& T_i^{~k}(P) e^i_{~(a)}(P)e_k^{~(b)}(P) =
T_a^{~b}(P)
\end{eqnarray}
are {\it physical or tetrad components} of the pseudotensor or
tensor field which describes an energy-momentum distribution, and $\bigl\{y^i\bigr\}$
are normal coordinates. $e^i_{~(a)}(y), e_k^{~(b)} (y)$ mean an
orthonormal tetrad $e^i_{~(a)}(P) = \delta_a^i$ and its dual $e_k^{~(a)}(P) = \delta_k^a $
paralelly propagated along geodesics through $P$ ($P$ is the origin
of the {\bf NRC(P)}).
We have
\begin{equation}
e^i_{~(a)}(y) e_i^{~(b)}(y) = \delta_a^b.
\end{equation}
For a sufficiently small 4-dimensional domain $\Omega$ which
surrounds {\bf P} we require
\begin{equation}
\int\limits_{\Omega}{y^i d\Omega} = 0, ~~\int\limits_{\Omega}{y^i
y^k d\Omega} = \delta^{ik} M,
\end{equation}
where
\begin{equation}
M = \int\limits_{\Omega}{(y^0)^2 d\Omega} =
\int\limits_{\Omega}{(y^1)^2 d\Omega} =
\int\limits_{\Omega}{(y^2)^2
d\Omega}=\int\limits_{\Omega}{(y^3)^2 d\Omega},
\end{equation}
is a common value of the moments of inertia of the domain $\Omega$
with respect to the subspaces $y^i = 0,~~(i= 0,1,2,3)$.
We can take as $\Omega$, e.g., a  sufficiently small analytic ball centered
at $P$:
\begin{equation}
(y^0)^2 + (y^1)^2 + (y^2)^2 + (y^3)^2 \leq R^2,
\end{equation}
which for an auxiliary positive-definite metric
\begin{equation}
h^{ik} := 2 v^i v^k - g^{ik},
\end{equation}
can be written in the form
\begin{equation}
h_{ik}y^i y^k \leq R^2.
\end{equation}
A fiducial observer {\bf O} is at rest at the beginning {\bf P}
of the used Riemann normal coordinates {\bf NRC(P)} and its four-
velocity is $v^i =\ast~ \delta^i_o.$ $=\ast$ means that an
equations is valid only in special coordinates.

We would like to note that we always will take $e^i_{~(o)} = v^i = \ast~\delta^i_o
$.

$\sigma(P;y)$ denotes the two-point {\it world function}
introduced in past by J.L. Synge \cite{Synge}
\begin{equation}
\sigma(P;y) =\ast {1\over 2}\bigl(y^{o^2} - y^{1^2} - y^{2^2}
-y^{3^2}\bigr).
\end{equation}
The world function $\sigma(P;y)$ can be defined covariantly by the
{\it eikonal-like equation} \cite{Synge}
\begin{equation}
g^{ik} \sigma_{,i} \sigma_{,k} = 2\sigma,
~~\sigma_{,i} := \partial_i\sigma,
\end{equation}
together with
\begin{equation}
\sigma(P;P) = 0, ~~\partial_i\sigma(P;P) = 0.
\end{equation}
The ball $\Omega$ can also be given by the inequality
\begin{equation}
h^{ik}\sigma_{,i} \sigma_{,k} \leq R^2.
\end{equation}
Tetrad components and normal components are equal at {\bf P}, so,
we will write the components of any quantity attached to {\bf P}
without tetrad brackets, e.g., we will write $S_a^{~b}(P)$
instead of $S_{(a)}^{~~~(b)}(P)$ and so on.

If $T_i^{~k}(y)$ are the components of an energy-momentum tensor
of matter, then we get from (5)
\begin{equation}
_m S_a^{~b}(P;v^l) = \bigl(2{\hat v}^l {\hat v}^m - {\hat g}^{lm}\bigr) \nabla_l \nabla_m {}
{\hat T}_a^{~b} = {\hat h}^{lm}\nabla_l \nabla_m {}{\hat T}_a^{~b}.
\end{equation}
Hat over a quantity denotes its value at {\bf P}, and $\nabla$
means covariant derivative.
Tensor $_m S_a^{~b}(P;v^l)$ is {\it the canonical superenergy tensor for matter}.

For gravitational field, substitution of the canonical
Einstein energy-momentum pseudotensor as $T_i^{~k}$ in (5) gives
\begin{equation}
_g S_a^{~b}(P;v^l) = {\hat h}^{lm} {\hat W}_a^{~b}{}_{lm},
\end{equation}
where
\begin{eqnarray}
{W_a^{~b}}{}_{lm}&=& {2\alpha\over 9}\bigl[B^b_{~alm} +
P^b_{~alm}\nonumber\cr
&-& {1\over 2}\delta^b_a R^{ijk}_{~~~m}\bigl(R_{ijkl} +
R_{ikjl}\bigr) + 2\delta_a^b{\beta}^2 E_{(l\vert g}{}E^g_{~\vert
m)}\nonumber\cr
&-& 3 {\beta}^2 E_{a(l\vert}{}E^b_{~\vert m)} + 2\beta
R^b_{~(a\vert g\vert l)}{}E^g_{~m}\bigr].
\end{eqnarray}
Here $\alpha = {c^4\over 16\pi G} = {1\over 2\beta}$, and
\begin{equation}
E_i^{~k} := T_i^{~k} - {1\over 2}\delta_i^k T
\end{equation}
is the modified energy-momentum tensor of matter \footnote{In
terms of $E_i^{~k}$ Einstein equations read $R_i^{~k} = \beta
E_i^{~k}$. If we admit $\Lambda$ term then we will have
$R_i^{~k} = \beta E_i^{~k} + \Lambda\delta_i^k.$}.

On the other hand
\begin{equation}
B^b_{~alm} := 2R^{bik}_{~~~(l\vert}{}R_{aik\vert m)}-{1\over
2}\delta_a^b{} R^{ijk}_{~~~l}{}R_{ijkm}
\end{equation}
are the components of the {\it Bel-Robinson tensor} ({\bf BRT}),
while
\begin{equation}
P^b_{~alm}:= 2R^{bik}_{~~~(l\vert}{}R_{aki\vert m)}-{1\over
2} \delta_a^b{}R^{jik}_{~~~l}{}R_{jkim}
\end{equation}
is the Bel-Robinson tensor with  ``transposed'' indices $(ik)$.
Tensor $_g S_a^{~b}(P;v^l)$ is the {\it canonical superenergy
tensor} for gravitational field $\bigl\{^i_{kl}\bigr\}$.
In vacuum $_g S_a^{~b}(P;v^l)$ takes the simpler form
\begin{equation}
_g S_a^{~b}(P;v^l) = {8\alpha\over 9} {\hat h}^{lm}\bigl({\hat
C}^{bik}_{~~~(l\vert}{}{\hat C}_{aik\vert m)} -{1\over
2}\delta_a^b {\hat C}^{i(kp)}_{~~~~~(l\vert}{}{\hat C}_{ikp\vert
m)}\bigr).
\end{equation}
Here $C^a_{~blm}$ denote components of the {\it Weyl tensor}.

Some remarks are in order:
\begin{enumerate}
\item In vacuum the quadratic form $_g S_a^{~b}{}v^a v_b$, where $v^av_a = 1$, is {\it
positive-definite} giving the gravitational {\it superenergy density} $\epsilon_g$
for a fiducial observer {\bf O}.
\item In general, the canonical superenergy tensors are uniquely
determined only along the world line of the observer {\bf O}. But
in special cases, e.g., in Schwarzschild spacetime or in Friedman
universes, when there exists a physically and geometrically
distinguished four-velocity $v^i(x)$, one can introduce in an
unique way the unambiguous fields $_g S_i^{~k}(x;v^l)$ and $_m
S_i^{~k}(x;v^l)$.
\item We have proposed in our previous papers to use the tensor $_g S_i^{~k}(P;v^l)$
as a substitute of the non-existing gravitational energy-momentum
tensor.
\item It can easily seen that the superenegy densities
$\epsilon_g := _g S_i^{~k}v^iv_k, ~~\epsilon_m := _m S_i^{~k}v^i v_k$
for an observer {\bf O} who has the four-velocity $v^i$ correspond
exactly to the {\it energy of acceleration} ${1\over 2}m {\vec a}{\vec a}$
which is fundamental in Appel's approach to classical mechanics
\cite{Appel}.
\end{enumerate}

In past we have used the canonical superenergy tensors $_g S_i^{~k}$
and $_m S_i^{~k}$ to local (and also, in some cases, to global)
analysis of well-known solutions to the Einstein equations like
Schwarzschild and Kerr solutions; Friedman and Goedel universes,
and Kasner and Bianchi I, II universes.
The obtained results were interesting (See \cite {Gar1}).

We have also studied the transformational rules for the canonical
superenrgy tensors under conformal rescalling of the metric
$g_{ik}(x)$\cite{Gar1,Gar2}.

The idea of the superenrgy tensors can be extended on angular
momentum also \cite{Gar1}. The obtained angular superenergy tensors
do not depend on a radius vector and they depend only on {\it
spinorial part} of the suitable gravitational angular momentum
pseudotensor \footnote{We have used in our investigation the
Bergmann-Thomson expression on angular momentum in general
relativity.}.

\section{Gravitational stability of the solutions to the Einstein equations and
canonical superenergy density}

By gravitational stability we mean stability of a background metric
${\tilde g}_{ik}(x)$ under small perturbations, see,
e.g.,\cite{LL,MD}
\begin{equation}
g_{ik}(x) = {\tilde g}_{ik}(x) + h_{ik}(x),
\end{equation}
where $\vert h_{ik}(x)\vert \ll\vert{\tilde g}_{ik}(x)\vert.$

Recently we have observed an exciting correlation between the total superenergy density,
 $\epsilon_s := \epsilon_m + \epsilon_g $, and gravitational stability of
 the solutions to the Einstein equations. Namely, we have noticed that
 when a solution is stable, then $\epsilon_s\geq 0$, and when the
 solution is  unstable, then $\epsilon_s <0$.
 \begin{center}
 The examples of the above mentioned correlation
 \end{center}
 \begin{enumerate}
 \item Exterior Schwarzschild with $\Lambda =0$ ------ stable ------- $\epsilon_s >0$:
 \item Einstein static universe ---- unstable ----- $\epsilon_s<0$;
 \item Kerr solution with $\Lambda =0$ --------------- stable ------ $\epsilon_s>0$;
\item Standard Friedman universes with $\Lambda =0$ ---------- stable ------ $\epsilon_s>0$;
\item Exterior Reissner-Nordstroem with $\Lambda =0$ -- stable ----- $\epsilon_s >0$;
 \item Minkowski spacetime --------- stable ----- $\epsilon_s =0$.
 \end{enumerate}

One can easily see that the above mentioned correlation follows
from the Proposition.

\vspace{0.1cm}
{\bf Proposition}

If the canonical total energy density $K_0^{~0}(y)$ has stable minimum at $P$, i.e., if $P$
is stability point of the analyzed solution, ${\tilde g}_{ik}(y)$, then $\epsilon_s(P) >0$.

\vspace{0.1cm}
{\bf Proof}. $\star$ Our proof lies on the constructive definition (5)
and on the following thermodynamical fact: a stable minimum of
the interior energy $U = U(S,V,N)$ is given by
\begin{equation}
\delta U = 0, ~\delta^2 U >0.
\end{equation}

We will apply the analogical conditions to the total canonical
energy density, matter and gravitation,
$_E K_0^{~0} \bigl(g; g^{ik};{}g^{ik}_{~~,l};{}, g^{ik}_{~~,lm}\bigr)$ in {\bf
NRC(P)}.\footnote{We use {\bf NRC(P)}in our proof but we write the
obtained results covariantly.} Namely, we put in {\bf NRC(P)}
\begin{equation}
\delta _E K_0^{~0}(P) =0, ~~\delta^2 _E K_0^{~0}(P) >0
\end{equation}
as conditions on stable minimum of the $_E K_0^{~0}(y)$ at the
point $P$.

Small metric perturbations (22) do not destroy such minimum
like as small variations $\delta S, ~\delta V, ~\delta N$ do not
destroy a local, stable minimum of $U$.

So, the local minimum  defined by (24) is a stable point of the
considered background solution ${\tilde g}_{ik}(y)$.

It is seen from (5), (8), (12) that the sign of the superenergy density
$S_a^{~b}(P) v^av_b =\star ~~S_0^{~0}(P)$
is determined by the sign of the integral in nominator of (5) because {\it (-) denominator
is always positive}.

Let $K_0^{~0}(y)$ has stable minimum at point $P$, ie., let $\delta _E K_0^{~0}(P) =0,
 ~\delta^2 _E K_0^{~0}(P)>0$. Then the analyzed solution is stable at this point.
One can see from (5),(8) and (12) that
$S_0^{~0}(P) = _g S_0^{~0}(P) + _m S_0^{~0}(P) = \epsilon_s (P)>0$ in the case because
$K_{(0)}^{~~~(0)}(y) - K_0^{~ 0} >0$ and the integral in nominator of (5) is positive.
 $\star$

\vspace{0.1cm}

From the Proposition it follows Conclusion that $S_0^{~0}(P)>0$
is {\it necessary condition} for gravitational stability in $\Omega$ (We
write this covariantly as $\epsilon_s (P) = S_i^{~k}(P) v^i v_k>0$.) $P\in\Omega$
is a running point of $\Omega$.

As the consequence of the Conclusion one has that if $S_0^{~0}(P)<0$
in the domain $\Omega$ ( We write this covariantly as $\epsilon_s(P) = S_i^{~k}(P)v^i v_k
<0$,  $P\in\Omega$.), then the considered solution cannot be gravitationally stable
in $\Omega$.

The stable flat Minkowskian spacetime gives an
example of a limiting case with $S_0^{~0} = \epsilon_s =0$.

\begin{center}
Some examples of the application of the above Conclusion
\end{center}
\begin{enumerate}
\item De Sitter spacetime ---- $\epsilon_s <0$ $\Longrightarrow$ The solution cannot
be gravitationally stable.
\item Anti-de Sitter universe ---- $\epsilon_s <0$ $\Longrightarrow$ The solution cannot
be gravitationally stable.
\item Bianchi I universe with $\Lambda =0$ ---- $\epsilon_s >0$ $\Longrightarrow$ This solution
can be gravitationally stable.
\item Kasner universe with $\Lambda = 0$ ---- $\epsilon_s>0$ $\Longrightarrow$ This solution
also can be gravitationally stable.
\item Expanding dust Friedman universes with $k=0,~~\Lambda<0$:
$\epsilon_s>0$ for small values of the cosmic time $t$, and $\epsilon_s<0$ for big values of
$t$. It means that the solution can be stable only for small
values of the cosmic time $t$.
\item Oscillating Friedman dust universes with $k=0, ~~\Lambda>0$:
$\epsilon_s >0$ for bigger values of the scale
factor $R(t)$ $\bigl(for~ t\in (\pi/3,{5\over 3}\pi)\bigr)$, and
$\epsilon_s<0$ for smaller values of $R(t)$ $\bigl( for~
t\in(0,{\pi\over 3};{5\over 3}\pi, 2\pi)\bigr)$. Thus, these
solutions can be stable only for bigger values of the scale factor
$R(t)$;
\item Expanding dust Friedman universe with $k=(-)1, ~~\Lambda<0$:
$\epsilon_s >0$ for small values of the cosmic
time $t$, and $\epsilon_s <0$ for big values of $t$. So, this
solution, likely as in the case 5., can be gravitationally stable
only for small values of $t$;
\item Exterior SdS static universe with $\Lambda >0$ and SadS static universe with $\Lambda<0$:
 $\epsilon_s<0$ for big values of the radial coordinate $r$, and $\epsilon_s >0$ for
small values of $r$. We conclude from this that these solutions
can be stable only for small values of $r$.
\end{enumerate}

Concerning more detailed information about the superenergy
densities cited above ---- see Apendix.
\vspace{0.1cm}

The our results concerning de Sitter and anti-de Sitter
universes seem to be supported by the recent papers
\cite{Garfin,Emel,Bizon}.

It is very interesting  that following our Conclusion the
gravitational stability of the considered dust Friedman models
with $\Lambda\not= 0$ depends on the evolutional phase of these
universes. It is sensible because $\Lambda<0$ gives here a
repulsive force which is growing with $t$ and, therefore, should produce instability,
and $\Lambda >0$ gives an additional attractive force growing with $R(t)$ and strenthening
gravitational stability.

In the first version of the paper (arXiv:1306.5121[gr-qc]) we have
conjectured that $\epsilon_s>0$ guarantees gravitational
stability. Now, we see that such conjecture was incorrect because $\epsilon_s >0$
does not ensure the local stability conditions (24). It only gives
{\it necessary condition} of stability.

\section{Final remarks}
On the {\it superenergy level} we have no problem with suitable tensor for
gravity, e.g., one can introduce gravitational {\it canonical superenergy
tensor}.
The canonical superenergy tensors, gravitation and matter, are useful to local analysis of
the solutions to the Einstein equations, especially to analyze
of their singularities \cite{Gar1}.

In this paper we have proposed a new application of these tensors  to study gravitational
stability of the solution to the Einstein equations.

The our proposal to study gravitational stability stability has thermodynamical origin
and it is different from approaches used already: an approach based on Lyapunov's stability
(connected with well-posed Cauchy problem) and dynamical system methods.

We think that this new application of the superenergy tensors can be useful.

\acknowledgments
This paper was mainly supported by Institute of Mathematics, University of Szczecin
(Grant No 503-4000-230351).

\section{Appendix}
We give here the canonical superenergy densities $\epsilon_s$ for
de Sitter, anti-de Sitter, static Einstein and Reissner-
Nordst\"om universes, for some dust Friedman universes with cosmological constant$\Lambda$
and for static SdS universe. For simplicity we will use in here the {\it
geometrized units} in which $G = c =1$.

As it was already mentioned we use the same notation and
definitions as in \cite{LL}, especially, the same form of the
Einstein equations without or with cosmological term, and the same form of
the FLRW line element.

The $\Lambda$ term we always treat as source term in Einstein
equations with energy-momentum tensor of the form
$_{\Lambda} T_i^{~k}  = (-) {\Lambda\over\beta}\delta_i^k.$

\begin{enumerate}
\item De Sitter spacetime ----- $\epsilon_s = (-) {28\over 27}\alpha \Lambda^2
<0$;
\item Anti-de Sitter spacetime ---- $\epsilon_s = (-) {32\over 27}\alpha\Lambda^2
<0$;
\item Einstein static universe ---- $\epsilon_s = (-){4\alpha\over
3R^4} <0$, where ${1\over R^2} = 4\pi\bigl(\rho +p\bigr)~= ~\Lambda - 8\pi p >0$;
\item Exterior Reissner-Nordstr\"om spacetime ----
\begin{eqnarray}
\epsilon_s &=& {2\alpha\over 9r^8}\bigl[3\bigl(2Q^2-r_s r\bigr)^2 +
5\bigl(Q^2 - r_s r\bigr)^2 + 2\bigl(3Q^2-r_s r\bigr)^2\nonumber\\
&+& 2\bigl(3Q^2 - r_sr\bigr)\bigl(2Q^2 - r_sr\bigr)\bigr]\nonumber\\
&+& {2Q^2\over r^8}\bigl(r_sr-2Q^2\bigr) + {12 Q^2\Lambda_{RN}\over
r^6}.
\end{eqnarray}
The last expression is positive for $r\geq r_H = m + \sqrt{m^2 -
Q^2}$, i.e., outside and on horizon $H$ of the Reissner-Nordstr\"om black hole.

Here $r_s := 2m,~~\Lambda_{RN} := 1 - {2m\over r} +{Q^2\over
r^2}$, and $m^2>Q^2, ~\alpha = {1\over 16\pi}, ~\beta = 8\pi$.
\item FLRW dust universes with $\Lambda\not= 0, ~k =0$.
In this case
\begin{equation}
\epsilon_s = {32\alpha\over 3}{{\ddot R}^2\over R^2} + {284\over
3}\alpha{{\dot R}^4\over R^4} - 124\alpha {{\ddot R}{}{\dot
R}^2\over R^3} + 12\alpha{{\dot R}{}{\dddot R}\over R^2}.
\end{equation}

For $\Lambda<0, ~k=0$  one has the solution of the suitable
Friedman equation \cite{Bondi}
\begin{equation}
R^3 = {3C\over 2\Lambda}\bigl[ch\{t(-3\Lambda)^{1/3}\}
-1\bigr], ~~C = {8\over 3}\pi\rho R^3 = const,
\end{equation}
from which it follows
\begin{equation}
 R(t) = A t^{2/3}
\end{equation}
for small $t$,
and
\begin{equation}
R(t) = B e^{bt},
\end{equation}
for big values of $t$.

Here $A,~B, ~b$ denote suitable, positive constants.

Substituting the asymptotic values  of $R(t)$ given by (28) and (29)
into (26), one gets
\begin{equation}
\epsilon_s = {9248\alpha\over 243 t^4}>0,
\end{equation}
for small $t$, and
\begin{equation}
\epsilon_s = (-) {20\over 3}\alpha b^4 <0,
\end{equation}
for big values of $t$.

For $\Lambda >0,~k=0$, one has the oscillatory solution to the
Friedman equation \cite{Bondi}
\begin{equation}
R(t) = A\bigl(1 - cosbt\bigr)^{1/3},
\end{equation}
where
\begin{equation}
A = \bigl({3C\over 2\Lambda}\bigr)^{1/3}, ~~b = (3\Lambda)^{1/3},
~~bt\in[0,2\pi].
\end{equation}
In this case the formula (26) gives
\begin{eqnarray}
\epsilon_s&=& {32\alpha\over 27} {b^4 \cos^2 bt\over(1-\cos bt)^2}
- {1148\alpha b^4\over 81} {\cos bt\sin^2bt\over(1-\cos
bt)^3}\nonumber\\
&+& {1232\alpha b^4\over 243}{\sin^4 bt\over(1-\cos bt)^4} -
{4\alpha b^4\over 3} {\sin^2  bt\over(1-\cos bt)^2}.
\end{eqnarray}

Again sign of the expression (34) depends on the evolutional phase
of this universe: for bigger values of $R(t)$, i.e., for
$t\in\bigl({\pi\over 3}, {5\over 3}\pi\bigr)$,
we have $\epsilon_s>0$, and
for smaller values of $R(t)$, i.e., for $t\in\bigl[(0,{\pi\over 3})\cup ({5\over 3}\pi,
  2\pi)\bigr]$ we have $\epsilon_s<0$.
\item Friedman dust universe with $\Lambda<0, ~k=(-)1$.

One gets in the case

\begin{eqnarray}
\epsilon_s &=& {32\alpha\over 3} {{\ddot R}^2\over R^2} -
{4\alpha\over 3R^2} - {280\alpha\over 3}{{\dot R}^2\over
R^4}\nonumber\\
&+& {284\alpha\over 3}{{\dot R}^4\over R^4} -124\alpha {{\dot
R}^2{\ddot R}\over R^3} + 12\alpha{{\dot R} {\dddot R}\over R^2} +
4\alpha{{\ddot R}\over R^3}.
\end{eqnarray}

Following Bondi \cite{Bondi} here we have

\begin{equation}
R(t) = A t^{2/3}
\end{equation}
for small values of $t$, and
\begin{equation}
R(t) = B e^{Dt},
\end{equation}
for big values of the cosmic time $t$.

Here $A, ~B, ~D$ mean suitable, positive constants.

Substituting the asymptotic values (36), (37) of the scale factor $R(t)$
into (35) one gets that
\begin{equation}
\epsilon_s = {\alpha\over 27 t^4} 1,688(8)>0,
\end{equation}
for small values of $t$, and
\begin{equation}
\epsilon_s = (-) {29\alpha\over 3}D^4<0,
\end{equation}
for big values of the cosmic time $t$.
\item Static SdS universe with $\Lambda >0$ and static SadS
universe with $\Lambda<0$.

In this case
\begin{eqnarray}
\epsilon_s &=& {12\alpha\over 9r^4}\bigl[{8\over 3}\bigl({r_s\over
2r} - {\Lambda r^2\over 3}\bigr)^2 + {1\over 4}\bigl(rr_s +
{\Lambda r^4\over 3}\bigr)^2\bigr]\nonumber\\
&+& {4\alpha\over 9}\bigl({r_s\over r^3} + {\Lambda\over
3}\bigr)^2 -{4\over 3}\alpha\Lambda^2,
\end{eqnarray}
where $r_s = 2m$.

It is easily seen from (40) that for big values of the radial
coordinate $r$ (We leave only the terms with $\Lambda$ in the
case)
\begin{equation}
\epsilon_s = (-) {20\alpha\over 27}\Lambda^2 <0,
\end{equation}
 and for small values of $r$ (We
omit here the terms with $\Lambda$)
\begin{equation}
\epsilon_s = {8\alpha\over 3} {r_s\over r^6}>0.
\end{equation}
\end{enumerate}

The total superenegy densities for the other solutions to the
Einstein equations mentioned in this paper have been  already
given in past \cite{Gar1}.

\end{document}